# Moving microfluidics ahead: Extending capabilities, accessibility, and applications

*Paul Blainey - Broad Institute / MIT*

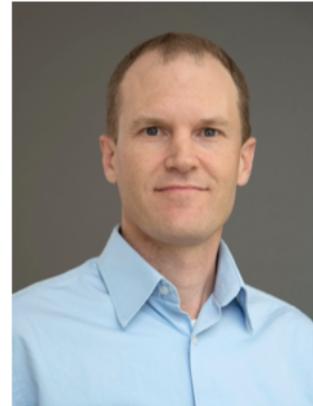

## Biography
*Dr. Blainey took degrees in mathematics and chemistry at the University of Washington before joining Professors Gregory L. Verdine and X. Sunney Xie in the Department of Chemistry and Chemical Biology at Harvard University. There, Dr. Blainey elucidated the mechanics of proteins diffusing along DNA using single-molecule mechano-optical assays and earned a PhD in Physical Chemistry. Dr. Blainey transitioned to Professor Stephen R. Quake's group at Stanford University, integrating active microfluidics, optical trapping, & new single-molecule microfluidic assays to enable single-cell microbial genomic sequencing. A faculty member in Biological Engineering at MIT and a Core Member of the Broad Institute since 2012, Dr. Blainey's group integrates microfluidic, molecular, and imaging tools to address the next generation of challenges in single-cell analysis, genomic screening, and therapeutics development.*

## Introduction
The field of microfluidics, which focuses on the control of fluids, solutes, and suspended particles on millimeter and smaller length scales, is undergoing rapid technical development and finding an expanding scope of applications. Here are presented three recently developed microfluidic technologies that extend the capability, accessibility, and applications of microfluidics.

## 1 – Integrated microfluidic sample preparation for genomic assays
Widespread clinical adoption of next-generation sequencing (NGS) is necessary to realize the full benefits of precision medicine, but currently, a "sample preparation bottleneck" makes NGS impractical for many clinical applications. Sample preparation is the series of processes that render biomass into a set of molecular sequence libraries. While improvements in sequencing instrument technology have driven down the cost of NGS, sample preparation has remained costly, complex, time-consuming, and error-prone. In particular, many clinical samples do not yield enough DNA, RNA, or chromatin for standard sample preparation methods. These challenges block full clinical deployment of NGS despite the last decade's advances in technology and genomic science.

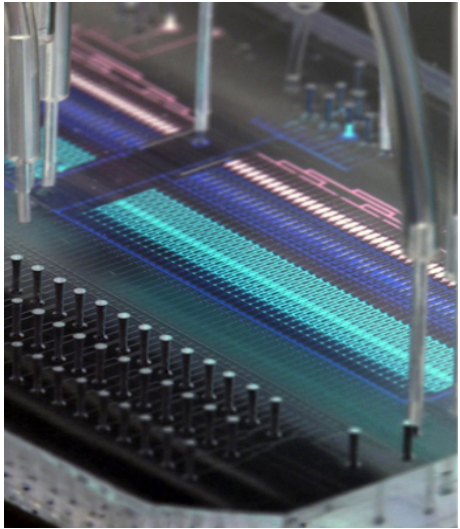

Addressing the sample preparation bottleneck with automated, versatile, microfluidic "lab-on-a-chip" devices -- thereby simplifying sample preparation for all clinically relevant genomic analyses (e.g. genome, epigenome, expression profile) -- is a longstanding goal in NGS technology development. Until now, technical challenges and market dynamics have prevented the integration of the entire sample preparation process within a single device. Recent advances in microfluidic instrumentation have focused on only a few steps in the sample preparation process, and are limited in their ability to meet the needs of high-demand clinical sequencing applications.

Microfluidics is well-suited for efficient sample preparation due to its capability for precise automated generation and manipulation of small fluid volumes.[1,2] Large sample batch sizes are needed to meet the incredible capacity of modern sequencing instrumentation, and parallel processing with microfluidics is a suitable way to fill this capacity. We constructed a two-layer microfluidic sample processing micro-architecture that enables batch sizes up to 96 samples in a single device smaller than a business card (see image, this section). Still larger batches are accessible by running multiple devices simultaneously. Our system integrates all the key sample preparation steps for genomic sequencing to increase throughput, reduce reagent consumption, reduce input requirements, reduce contamination, and improve reproducibility versus alternative manual or automated processes.

In our system, standard hydraulic micro-valves[3] partition on-device reactors for dead-end-fill loading[4] of precise quantities of each reagent which are actively mixed[5] in 36 nanoliter rotary reactors for DNA extraction and library construction operations. Each reactor operates in concert with a filter valve to strain cells or beads from solution as required for concentration and purification or pull-down operations. Filter valves are improved sieve valves[4] that are actuated to strain particles from solution and simplify the reaction circuit by allowing beads to be collected or released by flow in either direction. Excellent DNA extraction, reaction clean-up, and size selection performance is observed in the devices (better than 80% recovery of picogram-level starting material is typical for clean-up) using the solid phase reversible immobilization (SPRI) purification method.

The robust and efficient automated purification capability of our device allows "reset" of the sample in purified and concentrated form after each process step. This capability enables a paradigm shift in microfluidic sample processing systems, enabling (1) more direct translation of standard benchtop processes to the microscale and (2) the design of versatile micro-architectures such as our prototype devices that are capable of supporting a broad range of applications with arbitrary numbers of process steps. Our universal device design speeds development and validation of new applications by enabling effort to focus on protocol development and assessment rather than microfabrication of protocol-specific micro-architectural device variants.

We demonstrated the new microfluidic platform for microbial whole genome shotgun (WGS) sequencing with devices with capacities for 16-96 samples.[3] We showed DNA input reduction of more than 100-fold in an integrated cells-to-sequence library sample workflow while maintaining or improving variant calling accuracy for high GC organisms (clinical and environmental pseudomonas isolates) and difficult to lyse cells (M. tuberculosis and bacterial soil isolates). We also leveraged the enhanced throughput to sequence ~400 clinical Pseudomonas aeruginosa isolates and demonstrated Q68+ SNP detection performance that explained phenotypic antibiotic susceptibility observed across the isolates analyzed.

Altogether, our integration-focused approach and low-input capable purification technology will transform sample preparation to enable higher data quality to be obtained more quickly, more reliably, and at lower cost. Although advancement toward these objectives is broadly recognized as necessary in the field, our approach is different in substance and unique in its focus to solve the whole sample preparation problem for genomic research and clinical analyses.

## 2 – Hydrogel-based microfluidics for single-cell genome sequencing

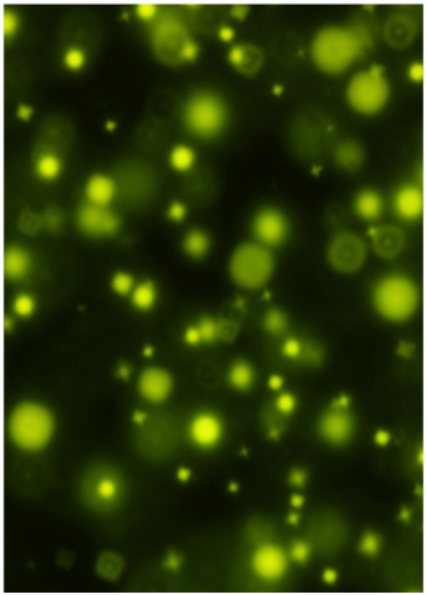

Microfluidic compartmentalization has been widely demonstrated using a variety of technical approaches including the lab-on-chip devices described above. However, cost, ease of use, and data quality are significant barriers to the broader deployment of microfluidic compartmentalization-based digital assays to meet the demand for single-molecule and single-cell analyses. Inspired by earlier work on polymerase cloning,[6] we developed a poly(ethylene glycol) (PEG) hydrogel system (see inset micrograph) as a general and facile method for compartmentalizing single molecules and single cells without discrete partitions in an approach we call virtual microfluidics.[7] Virtual microfluidics uses biocompatible hydrogels that can be functionalized to control their properties or anchor reagents, templates, or reaction products to the gel itself.

We demonstrate the potential of virtual microfluidics to support single-cell WGS by combining Escherichia coli and Staphylococcus aureus strains in a thin crosslinked PEG hydrogel. We lysed cells in situ by enzymatic and heat treatment, and supplied multiple displacement amplification (MDA) reagents to the immobilized template DNA. We recovered gel micro-samples manually and re-amplified the samples by MDA before constructing genomic sequence libraries and generating sequence data. Each micro-sample showed strong enrichment for reads mapping to Escherichia or Staphylococcus, but not mixtures of the two organisms, indicating that virtual microfluidics can resolve single cell amplification products from a mixed input.

On average, about 30% of the E. coli genome and about 60% of the S. aureus genome were covered in each single-cell sample, in-line with typical single microbe genome sequencing based on MDA. We found coverage uniformity in our single-cell punch samples to be comparable to previously published single-cell data sets based on liquid MDA with similar total amplification-fold. We also analyzed the occurrence of chimeric reads, which are known to occur with high frequency in MDA reactions by a cross-priming mechanism.[8] Chimeric reads directly confound de novo assembly, analysis of genome structure, and mapped read counting. Our single-cell datasets contained about 0.5% chimeric reads, approximately five-fold lower than previously published short-read datasets produced using standard single-cell MDA samples. This reduction in chimeric reads can be understood by restricted diffusion of the MDA reaction intermediates that prevents cross-priming that isolates each portion of the reaction mixture.

Virtual microfluidics enables high-throughput digital quantification of template cells or molecules and preparative single-cell whole-genome amplification without microfabricated consumables or expensive control systems. In addition, the straightforward addition of reagents to product clusters en mass and excellent optical access ideally suit the virtual microfluidics approach for in situ labeling of product clusters. We expect that virtual microfluidics will find application as a low-cost digital assay platform and as a high-throughput platform for single-cell sample preparation.

## 3 – Emulsion-based system for combinatorial drug screening

Biological networks are complex and nonlinear, yet we seek to perturb such networks to treat disease in a one-drug (one-target) paradigm that is fundamentally out of sync with our best understanding of biology. Our expanding knowledge of molecular networks and bio-active compounds represents a major opportunity to screen for combinations of compounds that co-target biological networks to achieve therapeutic effect. Drug combinations are of particular interest to improve efficacy and prevent the acquisition of drug resistance. In fact some existing drugs are known to be "dirty" in specificity terms (amiodarone) or to idiosyncratically engage multiple known targets (papaverine), the effects against whom in aggregate underlay observed therapeutic activity.[9]

Despite the promise of combinatorial therapies, little combinatorial screening is conducted due to the sheer scale of combinatorial space, which constitutes a blocking challenge across academia and industry. For example, a chemical library of just 1000 or 5000 compounds requires about 500,000 or 12.5 million independent tests respectively to cover all pairs of compounds, and 166 million or 20.3 billion independent tests to cover all unique compound trios. These scales challenge the limits of pharma-scale robotics and require a radically different approach.

We developed a prototype system for efficiently screening combinations of compounds in a micro-droplet emulsion format with optical readout. Our approach presents several advantages that together enable more efficient discovery of novel compound synergies. The current prototype eliminates complex specialized capital equipment (only a commodity microscope and emulsifier are needed), strongly suppresses compound crosstalk across droplets, requires less than 1/100[th] the quantity of compound needed by standard screening technologies, and runs in a large batch size for low cost and robust statistics. This new platform greatly expands the potential scale of combinatorial screening efforts and enables combinatorial screens that were not possible before.

Our prototype system (figure part A below) works by emulsifying each library compound with an RGB optical barcode (B) and the assay components (fluorescent bacteria and culture medium in our initial demonstration screen for anti-bacterial compound combinations). The resulting monodisperse emulsions (1 nanoliter aqueous droplets in a surrounding carrier oil) are then pooled to form a library of droplets that represents all the compounds to be screened. The droplet library is then introduced (A) to an array of microwells that capture pairs of droplets (B). We have optimized the protocol, reagent formulations, and microarray to prevent cross-talk of compounds across droplets, the major technical barrier that previously prevented the application of emulsion technology for drug screening.[10, 11] We image the array at 2x magnification to identify the compound in each droplet according to its optical barcode (B) and merge the droplet pairs simultaneously with an electric field (A). Here we use a parallel droplet processing approach rather than the serial processing approach that is most commonly applied in monodisperse droplet assays.[12] Parallel processing offers advantages in location-based indexing and robustness. For the P. aeruginosa growth assay shown, we incubate the array for 7 hours, and then image the merged droplet pairs to determine the extent of bacterial growth in the presence of each compound combination (C).

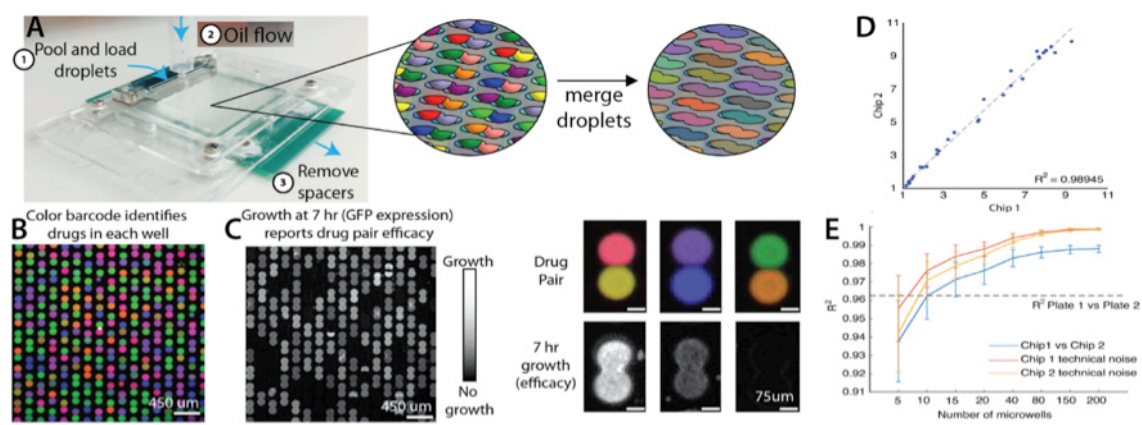

Our prototype system is capable of screening the 1800 pairwise combinations of 60 drugs on a single 60 mm x 60 mm silicone rubber chip containing 50,000 microwells (A). We set up this assay such that we observe each compound combination in an average of 10 micro-droplet assays in a single chip run, which provides an assay result at least as statistically robust as a single well in a standard 96 well plate assay (D,E). The level of micro-assay replication is continuously tunable to achieve, for example, $Z' > 0.5$ (high throughput screening standard for excellent quality) and $R^2 > 0.95$ between replicate chips as we have done for the bacterial growth assay (D,E). Our optical compound encoding strategy easily scales to hundreds of compounds to enable screening of a hundred thousand combinations or more per device.